\documentclass[useAMS,usenatbib]{mn2e}
\usepackage{graphicx}
\usepackage{amsmath}
\usepackage{natbib}
\usepackage{ulem}
\usepackage{amssymb}

\title[The Total Mass of Dark Matter Haloes]{The Total Mass of Dark Matter Haloes}
\author[D. Anderhalden and J. Diemand]{Donnino Anderhalden$^{1}$\thanks{E--mail:
donninoa@physik.uzh.ch} and J\"urg Diemand$^1$\thanks{E--mail:
diemand@physik.uzh.ch}\\
$^{1}$Institut f\"ur Theoretische Physik, University of Zurich, Winterthurerstrasse 190, 8057 Zurich}

\begin{document}

\def\mnras{MNRAS}
\def\apj{ApJ}
\def\apjs{ApJS}
\def\apjl{ApJL}
\def\araa{ARA\&A}
\def\aa{A\&A}
\def\physrep{Phys. Rep.}

\date{28 February 2011}

\pagerange{\pageref{firstpage}--\pageref{lastpage}} \pubyear{...}

\maketitle

\label{firstpage}
\begin{abstract}
The simple, conventional dark matter halo mass definitions commonly used in cosmological simulations ("virial" mass, FoF mass, $M_{50,100,200,\ldots}$) only capture part of the collapsed material and are therefore inconsistent with the halo mass concept used in analytical treatments of structure formation. Simulations have demonstrated that typical dark matter particle orbits extend out to about 90 per cent of their turnaround radius, which results in apocenter passages outside of the current "virial" radius on the first and also on the second orbit \cite[][]{Diemand2}. Here we describe how the formation history of haloes can be used to identify those particles which took part in the halo collapse, but are missed by conventional group-finders because of their remote present location. These particles are added to the part of the halo already identified by FoF. The corrected masses of dark haloes are significantly higher (the median mass increase is 25 per cent) and there is a considerable shift of the halo mass function towards the Press \& Schechter form. We conclude that meaningful quantitative comparisons between (semi-)analytic predictions of halo properties (e.g. mass functions, mass accretion rates, merger rates, spatial clustering, etc.) and simulation results will require using the same halo definition in both approaches.
\end{abstract}

\section{INTRODUCTION}\label{introduction}
Theoretical models of cosmological structure formation are based on the ansatz that an overdense region in the linearly evolved density field collapse into a dark matter halo when the linear density contrast reaches a certain collapse threshold $\delta_c$. In an idealised, radial spherical collapse, a region collapses into a point at the time when its linear density contrast grows to $\delta_c =1.686$ \cite[][]{Gunn}. \citet[][PS hereafter]{PS} have used this collapse threshold to calculate the abundance of haloes from the mass variance $\sigma^2(M)$ of the linear density field. PS was extended by \citet[][]{Bond} and many others using the 'excursion set' approach, which allows a more robust derivation of the original PS mass function and also the prediction of further halo properties like accretion rates, merger rates and spatial clustering \citep[see][for a recent review of this approach]{Zentner}. Note that in the PS formalism the mass enclosed in the linear over-density and the resulting halo mass are assumed to be equal, i.e. the {\it entire} collapsing region makes up the final mass of the halo.
\newline
\indent Since the collapse itself is fairly complicated (i.e. non-linear, clumpy, non-radial, non-spherical) event, theoretical models of structure formation cannot predict the detailed properties of the haloes they describe, i.e. PS does not describe (or depend on) the mass distribution in their haloes. Modern cosmological simulations on the other hand calculate the $z=0$ non-linear density field reliably, but to extract halo properties from simulations requires an operational halo definition, i.e. the choice and implementation of some kind of halo finder. Unfortunately the conventional halo definitions used by simulators are still based on an outdated, oversimplified picture: By imposing that a homogenous collapsing sphere reaches virial equilibrium, one finds that the material would settle within a sphere of half its turnaround radius, defined to be the virial radius $r_{\rm vir}$. By definition $r_{\rm vir}$ contains the virial mass $M_{\rm vir}$, which is assumed to be equal to the mass of the collapsed homogenous sphere. The moment of this virialisation is assumed to be when a radially collapsing sphere falls into one point. At this time the virial radius encloses $\Delta_M$ = 178 times the mean matter density in a flat universe with $\Omega_M = 1$. This simple picture motivates the operational definition of haloes as spherical overdensities (SO hereafter) with $\bar{\rho} = 178 \rho_{\rm crit}$ \citep[e.g.][]{Warren2}. A related halo definition is the friends-of-friends (FoF) algorithm \citep[e.g.][]{Davis}: FoF recursively links particles closer than some fraction $b$ of the mean particle spacing. The conventional value of $b=0.2$ produces groups with comparable mean density as the SO virial mass definition. The spherical collapse in $\Lambda$CDM leads to a larger $\Delta_M$ and therefore a smaller linking length $b$ \citep[see][for detailed comparisons of the halo mass function obtained with SO($\Delta_M$) and FoF($b$) for a wide ranges in $\Delta_M$ and $b$]{White1,White2,Warren,Tinker,Robertson}.
\newline
\indent SO and FoF are well defined and easy to use for the analysis of large data sets from cosmological N-body simulations. SO halo masses correlate rather tightly with some galaxy cluster observables (SZ, X-ray, optical), especially for large $\Delta_M$ values \citep[see][and references therein]{Tinker}. While FoF and especially SO based halo mass function are useful for comparisons with observations, they are not suited for comparisons with predictions from PS and related models, because they only capture a fraction of the collapsed mass: Simulations show that spheres which enclose the final $M_{\rm vir}$ of a halo, collapse only by about a factor of 1.4, i.e. much less than the factor of 2 assumed in the definition of the virial radius \citep[][]{Diemand}. Related indications that haloes are much more extended than their $r_{\rm vir}$ are their larger virialized regions \citep[][]{Maccio,Prada,Cuesta} and the fact that many haloes found between $r_{\rm vir}$  and 2$r_{\rm vir}$ of a large host halo are orbiting through and around this host, i.e. many were well inside its $r_{\rm vir}$ at some earlier time \citep[][]{Moore,Gill,Ludlow}. \citet[][]{Diemand2} have shown that the dynamics in outskirts of haloes are very well described by the self-similar secondary infall model from \citet[][]{Fillmore} and \citet[][]{Bertschinger}: Typical particle (and subhalo) orbits extend out to about 90 per cent of their turnaround radius and the apocenter distance decays only very slowly and asymptotes to about 83 per cent of the turnaround radius. This implies that after falling into a halo a typical particle has its first two apocenter passages beyond the current $r_{\rm vir}$ of this halo \citep[][]{Diemand2}. This causes significant amounts of is material which forms part of the collapsed region, and therefore of the halo mass in PS type models,  to lie beyond $r_{\rm vir}$. The mismatch between the PS mass concept and $M_{\rm vir}$ is obvious and implies that quantitative comparisons between models and simulations will require that both approaches agree on what exactly is meant by 'halo'.  
\newline
\indent One way out would be to combine analytic models with secondary infall to convert the total collapsed mass into a SO($\Delta_M$) mass for comparison with simulation results. A second solution is to maintain the PS halo mass concept and to develop a group-finder which is able to extract the entire collapsed halo mass from cosmological simulations. We have investigated this second route and present in this paper an algorithm which finds the total mass of haloes in the PS sense, based on a standard groupfinder (FoF is used in this work, although SO provides similar results) and on the formation histories of haloes given by their merger trees.
\newline
\indent This paper is structured as follows. After this overview of the mass problem, we describe in Section \ref{simulations} first the simulation setup we used for this work and thereafter the time dependent group finding algorithm. In Section \ref{MF} the resulting mass function based on this correction is presented whereas a summary and discussion is given in Section \ref{discussion}.

\section[]{THE CODE}\label{simulations}
In this section we present our correction code which computes the total halo mass. Sorting as well as searching algorithms are taken from {\it Numerical Recipes} \cite[][]{NR}.
\subsection{N-body Simulations}
For this work we make use of two different $\Lambda$CDM simulations, which were run using the parallel tree code PKDGRAV \citep[][]{Stadel}. Both runs have a total number of $512^{3}$ particles, simulated in cubic volumes of $80 h^{-1}$ Mpc (Run 1) and $250 h^{-1}$ Mpc (Run 2) respectively, leading to particle masses of $3.18\times10^{8} h^{-1}$ M$_{\odot}$ and $9.67\times10^{9} h^{-1}$ M$_{\odot}$. This setup enables us to probe a mass range of almost five orders of magnitude, although the statistics at the high mass end is not sufficient to give a detailed prediction for the cluster mass function. We are using a WMAP 1$^{st}$ year cosmology for both simulations. The parameters are
\begin{eqnarray}\label{parameters}
p & = & (\Omega_0, \Omega_{dm}, \Omega_{\Lambda}, h, \sigma_8, n) \nonumber \\
 & = & (0.3, 0.255, 0.7, 0.7, 0.9, 1).
\end{eqnarray}
The initial conditions for both simulations were created with GRAFIC2 \citep[][]{Bertschinger2}; the analysis described in the following is based on 40 snapshots in time, ending at the present epoch.

\subsection{Finding the Total Halo Mass}\label{code}
According to the PS ansatz, the mass of a dark matter halo is defined to be the entire mass which once collapsed from a connected (spherical or ellipsoidal) and overdense region. In a first attempt to find this mass, all particles in haloes today were traced back to their initial positions. Then we tried to find a well defined way to complete the initial particle distribution into a convex and connected form. Unfortunately, these primordial regions turned out to be highly irregular and they are often not connected, so it was not possible to come up with a method to identify all particles which will collapse into a halo using only the initial distribution of those particles that ended up in haloes identified by the FoF groupfinder. We therefore consider the entire formation history of each halo in order to compute the total collapsed mass. The goal is to detect all particles which did belong to a progenitor of a redshift $z=0$ FoF-group\footnote{We focus on $z=0$. The same method can be applied at higher redshifts, if enough earlier snapshots are stored and analysed.} and to join them together with the part of the halo already found by FoF into one complete $z=0$ halo.

\begin{figure}
\begin{center}
\includegraphics[scale=0.28]{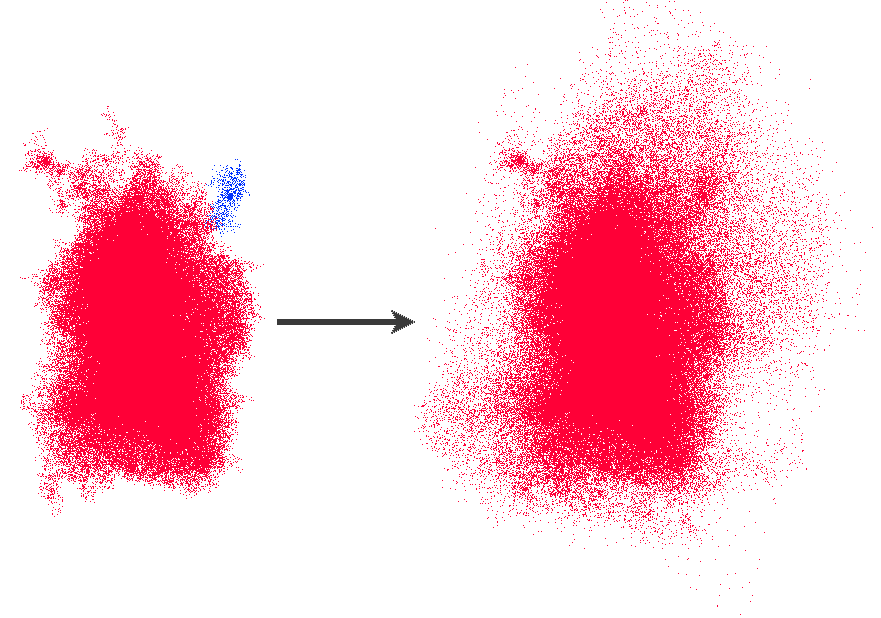}
\caption{Projection on the x-z-plane of a cluster-sized halo at $z=0$. The left panel shows all particles which are members of the clusters $z=0$ FoF group (red), and the particles in a smaller, nearby FoF group (blue). After correction for lost particles (right panel), many more particles, including the small FoF group now form part of the total mass of the cluster. The trajectory of the small group reveals that it is actually a subhalo (its irregular shape is due to tidal stripping, some of its tidal stream is captured by its FoF group).}
\label{tipsy}
\end{center}
\end{figure}

First, a halo merger tree is constructed based on the FoF group snapshots with a linking length $b=0.2$. This is done by comparing the number of particles which went from a progenitor group $G_{z_i}$ to a group $G_0$ at present time ($\mathcal{N}_{G_{z_i} \rightarrow G_0}$, the subscript $0$ denotes redshift $z=0$), with the number of particles which were in the progenitor $G_{z_i}$,
\begin{eqnarray}
\mathcal{R} \equiv \frac{\mathcal{N}_{G_{z_i} \rightarrow G_0}}{N(G_{z_i})},
\end{eqnarray}
and demanding that $\mathcal{R} \geq 0.5$ $\forall z_i$ to ensure unique remnants and merger tree without splits.

Now the total collapsed mass for each object is found by going through all particles and assigning each one to the correct $z=0$ halo. If a particle was part of a progenitor, the code assigns it to the unique successor according to the merger tree. Particles which do not already form part of the correct successor FoF group are added to that halo. Most of the corrected particles do not form part of any $z=0$ FoF group, i.e. they have orbited through a progenitor halo and are now found somewhere in the low density outskirts of the $z=0$ successor FoF group. Some of the corrected particles are members of a different $z=0$ FoF group, i.e. they form part of a subhalo which did fly through a progenitor halo and is now found outside the extent of the $z=0$ successor FoF group. In a conventional analysis such a subhalo would be misclassified and counted as an individual field halo. Our approach realises that this halo used to be a subhalo and should still be counted as a satellite, its mass is added to the primary halo and the subhalo disappears from the halo mass function\footnote{A particle can only be part of one halo, i.e. in our corrected mass function no mass element can be counted more than once.}. Note that such extended, but bound orbits are typical for particles and subhaloes which are on their first two orbits after infall \citep[][]{Diemand2}. 

\begin{figure}
\begin{center}
		\includegraphics[scale=0.43]{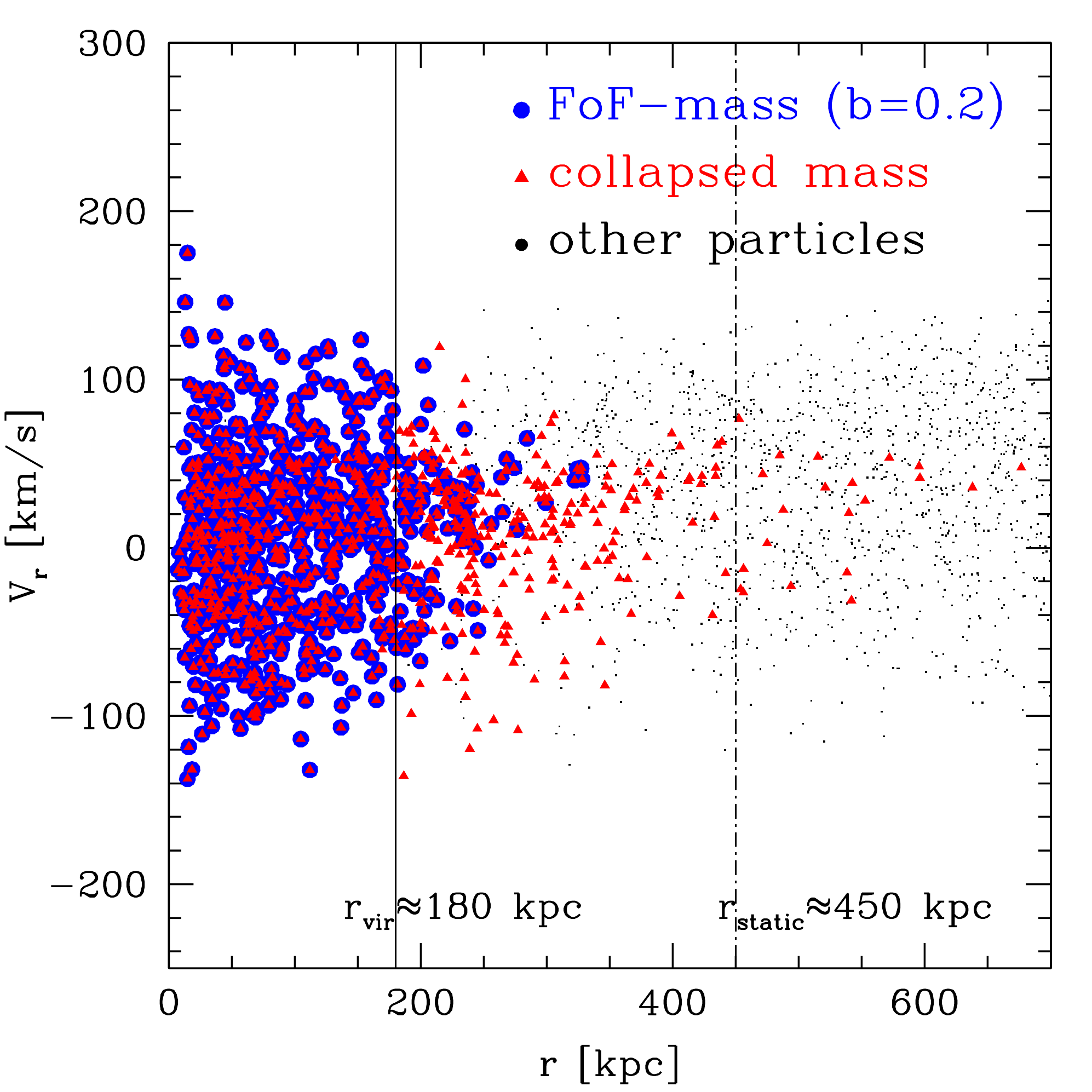}
		\caption{Phase-space distribution of dark matter particles at redshift $z=0$ of a small galaxy-sized halo. Particles in the FoF group are plotted as blue circles, red triangles give all halo members after the correction. The black dots are particles which are neither in the FoF group, nor in the corrected halo. In addition to the virial radius (solid line), the static radius (dot-dashed) is given \citep[][]{Cuesta}. Depending on the definition, the mass of this halo is $M_{\text{vir}} = 2.13\times10^{11}$ M$_{\odot}$, $M_{\text{FoF}} = 2.37\times10^{11}$ M$_{\odot}$, $M_{\text{static}} = 4.73\times10^{11}$ M$_{\odot}$ or $M_{\text{total}} = 3.26\times10^{11}$ M$_{\odot}$.
		\label{phasespace}}
\end{center}	
\end{figure}

An illustration of one example outcome of our correction method is shown in Fig. \ref{tipsy}. In the FoF result (left panel) there are two distinct haloes, a cluster-sized halo (red, $M_{\rm FoF} = 1.16 \times10^{14}$ M$_{\odot}$) and a nearby smaller halo (blue, $M \sim10^{11}$ M$_{\odot}$). Following the orbit of the small halo back in time shows that this halo is orbiting around the cluster and that it did form part of the clusters progenitors at earlier epochs, i.e. it is actually a subhalo which happens to lie outside its $z=0$ primary. Therefore our correction also takes care of all these subhaloes and adds them to their hosts in addition to the many other particles which were lost and not captured in the hosts $z=0$ FoF group.
\newline
\indent Fig. \ref{phasespace} shows the phase-space distribution of an FoF group and of the corresponding complete, corrected halo. The bulk (almost 90 per cent) of the additional particles from the correction are located between one and two virial radii. However, about 5 per cent of the corrected particles have travelled out to a distance of more than three virial radii after their infall. These exceptional distances suggest that these particles have gained kinetic energy during their passage through the halo, due to real \citep[e.g. three body encounters involving a massive subhalo,][]{Sales} or numerical effects. Some of these very remote particles might never fall back into this halo. At $z=0$ they might be unbound or even bound to another halo. Nevertheless, for the total halo mass and especially for the purpose of comparison with analytic results, we advocate counting even those remote particles as halo members, since they did fall in once and PS and related approaches do not account for the complicated nonlinear effects which led to their ejection.\footnote{In a few cases two nearby $z=0$ FoF groups, which according to our merger tree do not appear to have ever been one and the same halo (i.e. merged and de-merged), might still have exchanged a significant number of particles. If the correction assigns more than 15 per cent of the members of the smaller group to its larger neighbour we assume that the small group is a subhalo and add it to its primary halo. Such haloes did have a close encounter with the main halo, but they were not caught inside the main halo at the analysed snapshots. An overlap of less than 15 per cent is assumed to be due to ejected particles and the two haloes are kept in the catalogue as individual objects. The threshold value of 15 per cent is arbitrary. However, we have checked that our resulting mass function is practically unaffected by this threshold value, as long as it is larger than a few percent.}

\begin{figure}
		\includegraphics[scale=0.43]{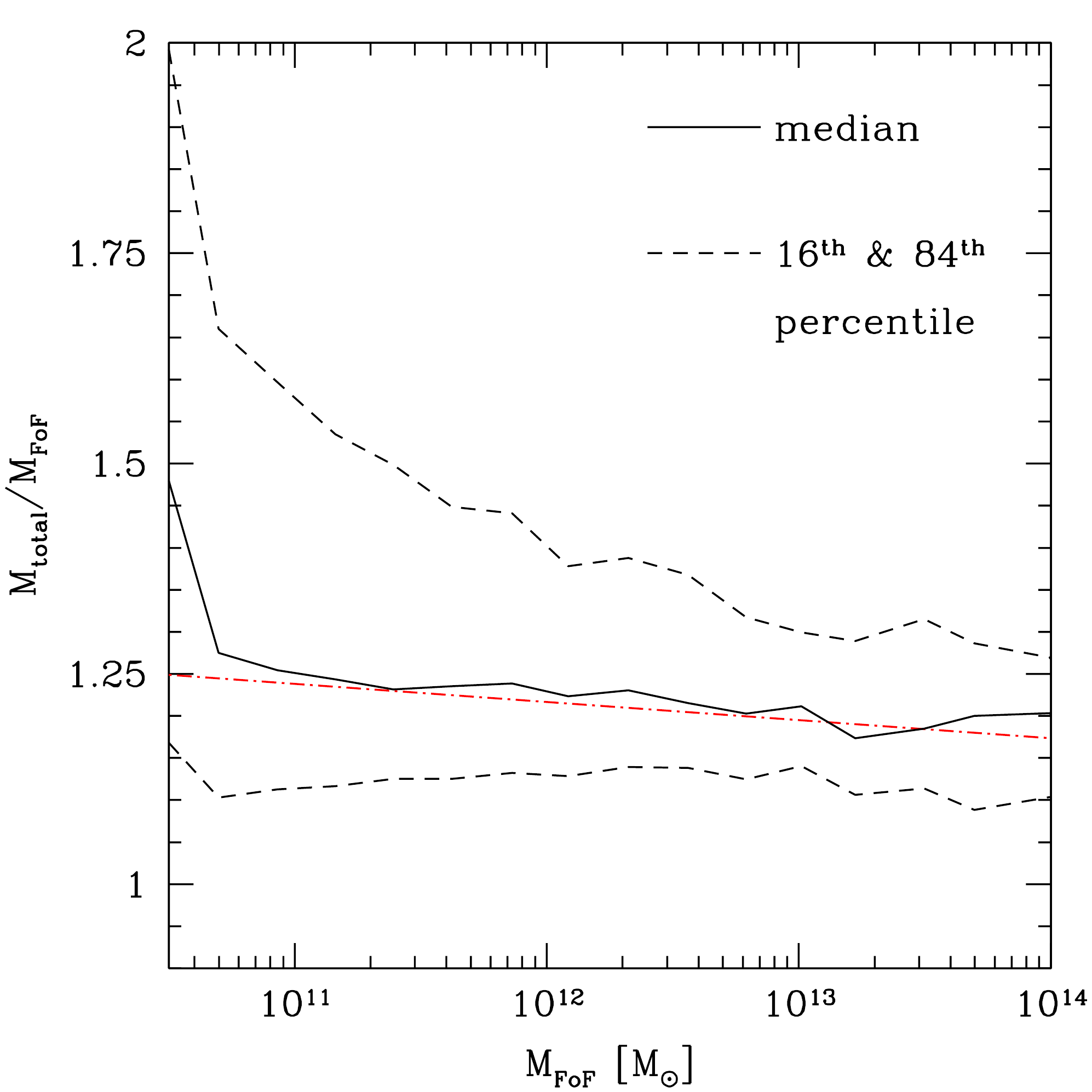}
		\caption{
		The total collapsed mass-to-FoF mass relation at $z=0$. The solid line shows the median, the dot-dashed lines the $16^{\text{th}}$ and $84^{\text{th}}$ percentile. The red dot-dashed line is a linear fit to the median: $M_{\text{total}}/M_{\text{FoF}} = 1.25 - 9\times10^{-3}\log(M_{\text{FoF}}/3.16\times10^{10}$M$_{\odot}$).}
		\label{median}
\end{figure}

Fig. \ref{median} shows the {\it total collapsed mass-to-FoF mass} relation for all dark matter haloes at redshift $z=0$. For clarity, only the median as well as the $16^{\text{th}}$ and $84^{\text{th}}$ percentiles are plotted. At the low mass end the scatter is largest: some haloes are up to three times heavier than their corresponding FoF group, while other haloes have about the same mass before and after the correction. The median correction values are almost independent of mass, $\sim$25 per cent for galaxy-sized haloes and $\sim$20 per cent for clusters. The mass dependence is weak, because the secondary infall pattern with subsequent apocenter passages beyond the virial radius applies to both galaxy- and cluster-sized haloes \citep[][]{Diemand2}. Only the weighting of these patterns differs somewhat: in clusters a larger mass fraction is currently infalling and the relative importance of the older, distant material near its first and second apocenter is slightly smaller.
\newline
\indent The static mass \citep[][]{Cuesta} is the mass of all particles enclosed in the static radius. By definition, the static radius is the largest radius where the mean radial velocity averaged over all haloes in this mass bin is still close to zero, i.e. $< 0.05 v_{\text circ} (r_{\text vir})$. For the galaxy-sized halo in Fig. 2, the static mass is about 40 per cent larger than the our total mass. This originates from the fact that $M_{\text{static}}$ encloses all particles up to $r_{\text{static}}$ and many of those have never experienced infall (black dots in Fig. \ref{phasespace}). Another difference between $M_{\text{static}}$ and $M_{\text{total}}$ is that $M_{\text{static}}$ includes all particles within an increased search radius around each halo: If two FoF groups lie close together they do both prevail and grow in mass, and some of the additional mass may be assigned to both haloes. In the case of $M_{\text{total}}$ every particle can contribute to at most one halo. If two nearby FoF groups emerge from a common progenitor, they are joined together into one larger halo. For galaxy-sized haloes our total-to-FoF mass ratios are smaller than the typical static-to-virial mass ratio in \citet[][]{Cuesta}, whereas it is the other way around for clusters. Our correction is slowly decreasing with mass (see linear fit in Fig. \ref{median}), while the static-to-virial mass ratio is an increasing function at the low mass end. Above the characteristic mass ($M_{\star}$) the typical static-to-virial mass ratio decreases quickly because the relative importance of infall increases, which pushes the static radius back close to the virial radius. Our method shows that haloes well above $M_{\star}$, e.g. cluster haloes at $z=0$, still do have a total mass which is significantly larger than their FoF and their virial mass. Below $M_{\star}$ mass corrections (up to 2-3 times the FoF mass) are more common, causing the 84$^{\text{th}}$ percentile to increase.

\section{THE MASS FUNCTION}\label{MF}
\subsection{Correcting the Mass Function}
In this Section we present the halo mass function (MF hereafter) based on the total halo mass, instead of the conventional choice of FoF or SO halo masses. The PS \citep[][]{PS} result, as outlined in Section \ref{introduction}, is given by
\begin{eqnarray}\label{massfunctionrep1}
n(M,t) dM = \frac{\bar{\rho}}{M^2} f_{PS} \frac{d \log \nu}{d \log M}dM,
\end{eqnarray}
where $\bar{\rho}$ is the mean background density of the universe, $\nu \equiv \frac{\delta_c(t)}{\sigma(M)}$ and
\begin{eqnarray}
\label{press}
f_{PS}(\sigma) = \sqrt{\frac{2}{\pi}} \nu \exp\Big(-\frac{ \nu^2}{2}\Big)
\end{eqnarray}
is the multiplicity function which gives the fraction of mass associated with haloes in a unit range of $\log \nu$. \citet[][ST hereafter]{ST} proposed a more general formula motivated by the assumption of an ellipsoidal collapse,
\begin{eqnarray}
\label{sheth}
f_{ST}(\sigma) = A \sqrt{\frac{2a}{\pi}}\Big[ 1 + \big(\frac{1}{a \nu^{2}} \big)^p \Big] \nu \exp\Big(-\frac{a \nu^2}{2}\Big),
\end{eqnarray}
where $A=0.322$, $a=0.707$ and $p=0.3$ \citep[see][for a discussion of elliptical collapse and collapse barriers]{Desjacques,Robertson}. In the limit of $A=\frac{1}{2}$, $a=1$ and $p=0$ it reduces to the spherical collapse and to the PS form. Recent work by \citet[][]{Bhatta} suggests a new fitting function, similar to the ST form, but with one additional parameter. Using 67 high resolution simulations, they found that ST deviates up to 40 per cent from their simulated FoF mass functions at the high mass end (Fig. 4 in \citet[][]{Bhatta}). At redshift $z=0$ their MF takes the form
\begin{eqnarray}
\label{bhattafit}
f_{mod}^{ST}(\sigma) &=& 0.333\sqrt{\frac{2}{\pi}}\exp\Big(-\frac{0.788 \delta_c^2}{2 \sigma^2}\Big)\times \\ \nonumber
& & \Big[1 + \Big(\frac{\sigma^2}{0.788 \delta^2_c} \Big)^{0.807} \Big]\Big(\frac{\delta_c \sqrt{0.788}}{\sigma}\Big)^{1.795}.
\end{eqnarray}

\begin{figure}
\centering
\includegraphics[scale=0.43]{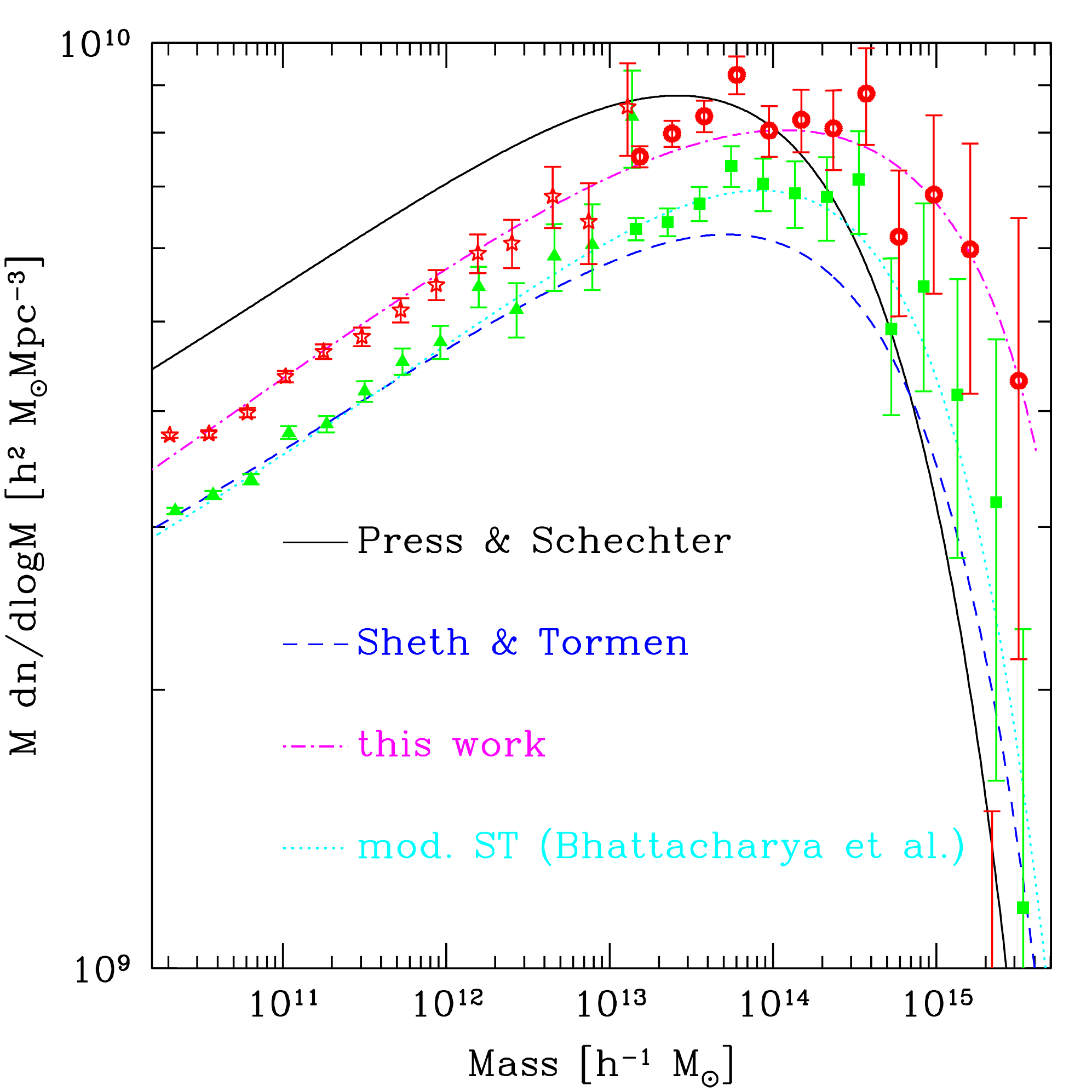}
\caption{Halo mass functions extracted at $z=0$ form our two simulation boxes, using the conventional FoF ($b=0.2$) halo mass definition (green triangles and squares corresponding to Run 1 and Run 2 resp.) and using the corrected total halo masses (red stars and circles corresponding to Run 1 and Run 2 resp.). The errorbars represent the Poissonian error. For comparison the PS, ST and \citet[][]{Bhatta} mass functions (black solid, blue dashed and cyan dotted lines) are plotted.
The purple, dot-dashed line shows a fit (see Eq. \ref{fittingfunction}) to our corrected mass function.
\label{massfunction}}
\end{figure}
The FoF mass functions from our two simulations, with a linking length $b=0.2$, is plotted with green data points in Fig. \ref{massfunction}. Only haloes with more than 50 particles are considered, yielding a lower mass limit of $2.3\times10^{10}$ M$_{\odot}$. At the low mass end, the ST form of Eq. \ref{sheth} provides a good fit, but it significantly starts to deviate at high masses. The recently proposed modified ST fitting formula of Eq. \ref{bhattafit} fits our FoF based mass function very well over the entire mass range probed by our two simulations.
\newline
\indent \citet[][]{Bhatta} as well as other empirical fitting functions \citep[e.g.][]{Jenkins,Warren,Tinker,Reed} are based on conventional mass definitions like $M_{\text{vir}}$, $M_{\text{FoF}}$, $M_{50,100,200,\ldots}$, which capture only a fraction of the total collapsed mass. They are therefore not suited for comparisons with analytical models of the PS and ST type. The corrected mass function (CMF hereafter), using the total halo masses after our correction, is plotted as red stars and circles in Fig. \ref{massfunction}. Here the errorbars represent the Poissonian error. Below $M_{\star}$ there is a shift of the mass function of about 15 per cent relative to the one based on FoF halo masses. In the cluster regime this shift seems to increase, but in order to get more reliable predictions a larger box simulation is needed.
\newline
\indent Based on the minimal-parameter multiplicity function described in \citet[][]{Warren} and by using the Levenberg-Marquardt algorithm, we determine a best fit for the CMF of the form:
\begin{eqnarray}
\label{fittingfunction}
\hspace{-0.6cm}& &\hspace{-0.6cm} f_{CMF}(\sigma,z=0) = a(\sigma^{b} + c)\exp\Big(-\frac{d}{\sigma^2}\Big),\nonumber \\
\hspace{-0.6cm}& &\hspace{-0.6cm} a = 2.825, \quad b = 0.138, \quad c = -0.883, \quad d = 0.154. \nonumber \\
\end{eqnarray}
This fitting formula is represented by the purple line in Fig. \ref{massfunction}.

\subsection{Comparison with Higher Linking Lengths}

\begin{figure}
\centering
\includegraphics[scale=0.43]{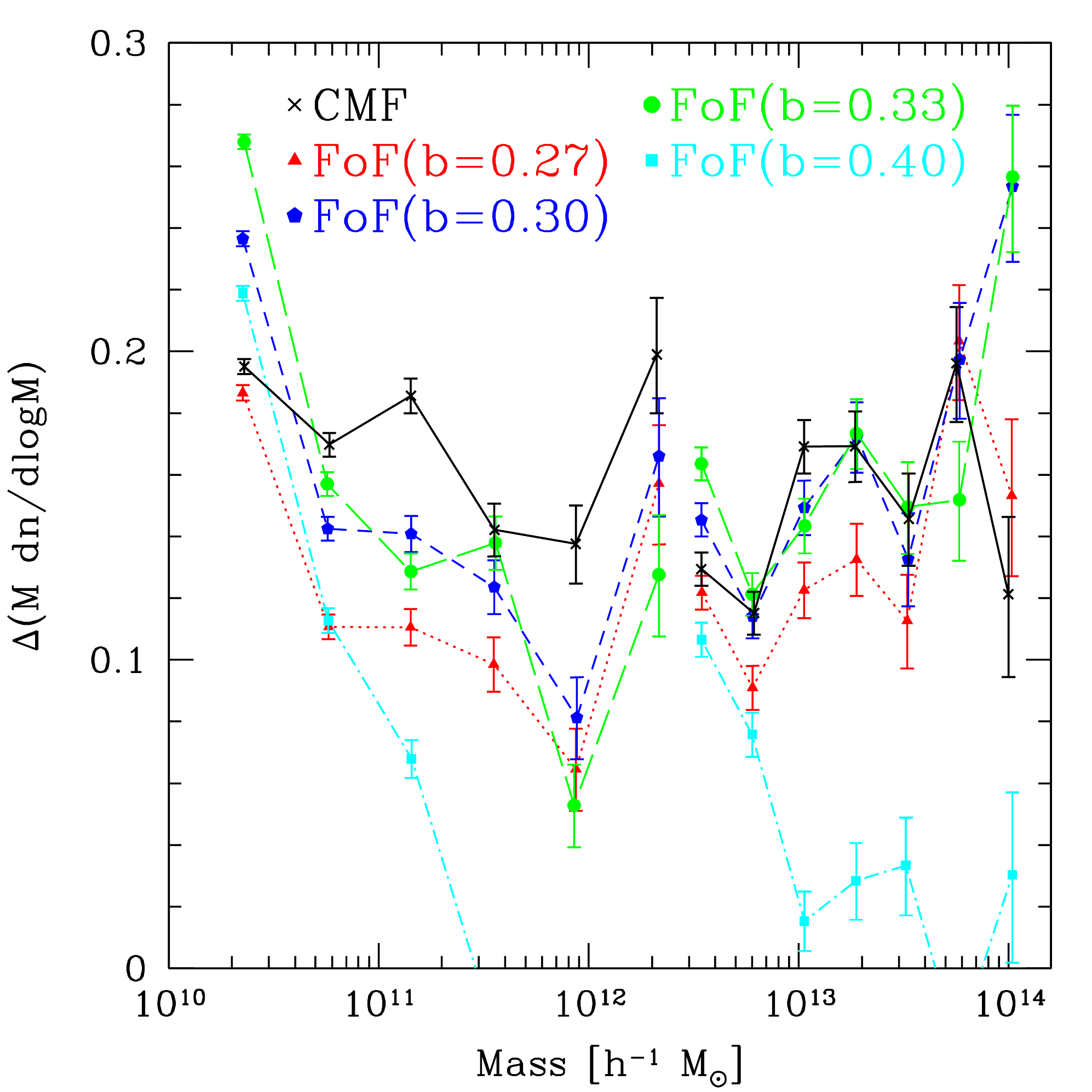}
\caption{Residuals between the measured FoF($b=0.2$) mass function and the CMF (solid line) at $z=0$. In addition also the residuals between the measured mass functions analysed with varying linking length parameter, $b=0.27, 0.30, 0.33$ and 0.40 (dotted, dashed, long-dashed and dot-dashed), are shown. The left and right data sets correspond to Run 1 and Run 2 respectively.
\label{fofratio}}
\end{figure}

As shown in Fig. \ref{median} the total collapsed mass for individual haloes lies approximately 20 to 25 per cent above the the FoF($b=0.2$) mass. This is also reflected in the representation of the mass function in Fig. \ref{massfunction}. A simple way to increase the mass of a halo is to increase the linking length parameter in the FoF algorithm (or similarly relax the virial overdensity criterion in an SO group finder). In this subsection we show that the total collapsed mass and therefore the CMF cannot be reconstructed by simply varying the only free parameter in the FoF group finder. Fig. \ref{fofratio} shows the effects of increasing the FoF linking length and compares them with our CMF.
None of the FoF mass functions with increased linking length are able to reproduce the CMF, even though the deviations between the CMF and FoF($b=0.27-0.33$) mass function tend to get smaller above $M_{\star}$.

For an even more unrealistic value of the linking length, $b=0.40$, the deviations get quite large. In this case, other unphysical characteristics like interconnections of groups through slightly overdense filaments start to dominate and therefore changing the overall shape of the mass function due a decrease of the the total number of galaxy-sized objects and the sudden occurrence of super-clusters of order $10^{16-17}$M$_{\odot}$.

\subsection{Universality of the Corrected Mass Function}
The extended PS formalism is from its analytic arguments expected to be universal, i.e. its predicted halo abundance depends only on the variance of the density field $\sigma(M)$, but not on redshift or cosmology \citep[e.g.][]{Zentner}.  However, it has been shown by various authors that universality only holds at the 10 to 20 per cent level when using FoF mass functions and slightly worse if haloes are identified with an SO group finder \citep[e.g.][]{Lukic,Tinker,Bhatta}.

To check if the CMF follows the universal behaviour, we ran two additional simulations with the same cosmology as in Eq. \ref{parameters}, but with a significantly higher initial redshift, and box sizes of $40 h^{-1}$Mpc and $80 h^{-1}$Mpc respectively. By keeping the number of particles constant, these simulations provide a sufficiently high resolution at redshifts $z=1,2,3$ not only to recover the FoF mass function, but also to apply our correction algorithm (i.e. tracing back particles over several time steps). For testing universality it is preferable to represent the mass function in the $f(\sigma) - \ln\sigma^{-1}$ plane,
\begin{eqnarray}\label{massfunctionrep2}
f(\sigma) = \frac{M}{\rho_0}\frac{dn}{d\ln \sigma^{-1}},
\end{eqnarray}
since cosmology and redshift dependences are absorbed in $\sigma(M)$.
\begin{figure}
\centering
\includegraphics[scale=0.43]{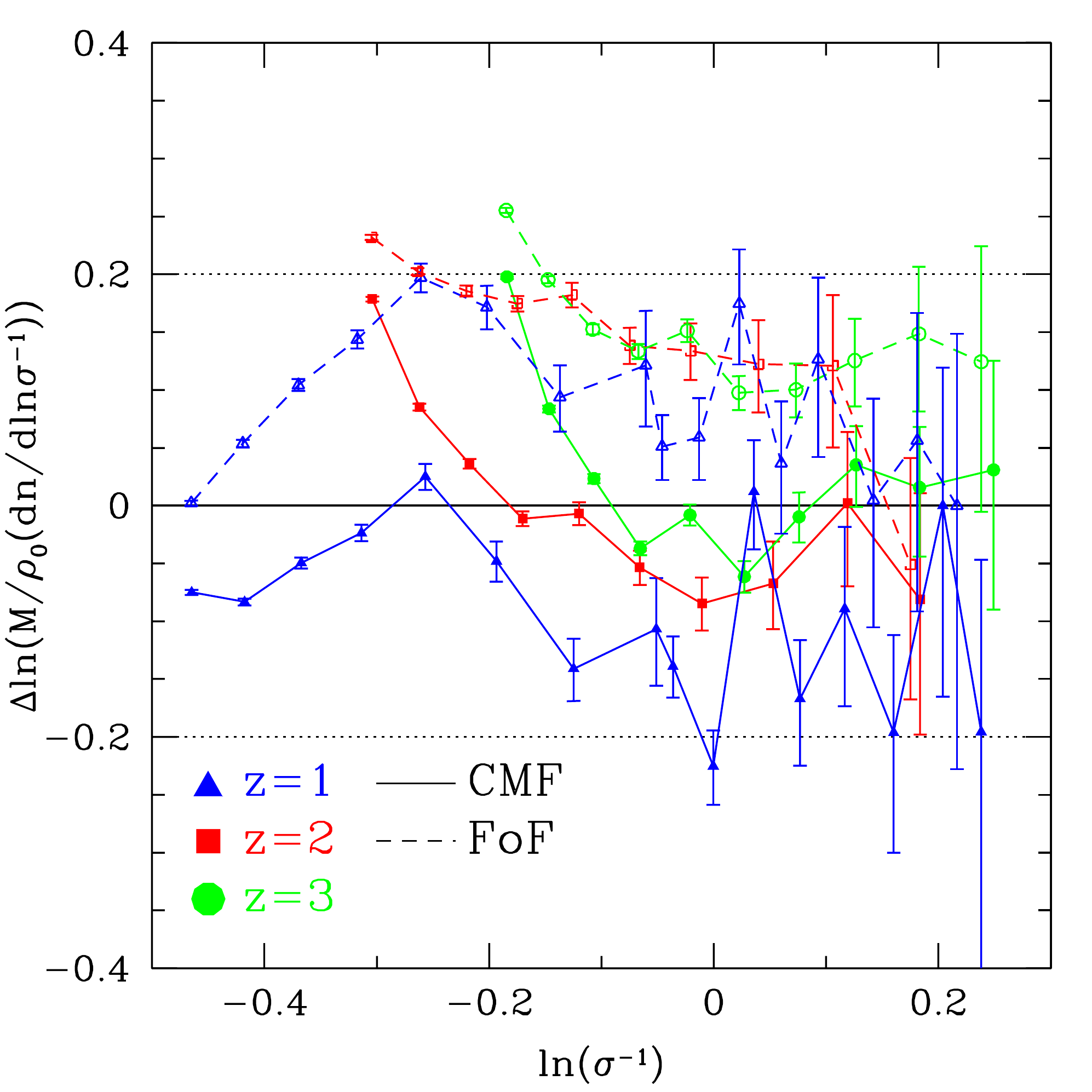}
\caption{Deviations from universality of the measured FoF (dashed lines) and CMF mass function (solid lines) at redshifts $z=1$ (blue), $z=2$ (red) and $z=3$ (green). The horizontal lines represent the 20 per cent threshold often found in the literature. Correcting the halo mass as described in Section \ref{code} shifts the mass function towards universality.
\label{residuals}}
\end{figure}

In Fig. \ref{residuals} the deviations from universality of the measured FoF and CMF mass functions at redshifts $z=1,2$ and 3 are shown. Whereas the FoF mass function is universal approximately at the 15 per cent level up to redshift $z=3$ (consistent with previous studies \citep[e.g.][and references therein]{Bhatta}), we notice a significant shift towards universality in the CMF case for redshifts $z \gtrsim 1$. This should not be surprising since the goal of the CMF was to approach the halo mass definition in the sense of PS as close as possible.

\citet[][]{Courtin} investigated the behaviour of mass functions in varying dark energy models. They found that the virialization process contributes to shaping the mass function in a cosmological and redshift dependent way and therefore a clear break down of universality when dark energy starts to dominate ($z\lesssim1$). In the case of the total collapsed mass, the virialization process is made more independent of redshift and hence the CMF is expected to show a higher degree of universality.

\section{SUMMARY AND DISCUSSION}\label{discussion}
In this letter we present a group finding algorithm which, based on a conventional FoF group finder, identifies the total collapsed mass of every halo in the sense of the PS theory and its extensions. We found a considerable shift of the mass function of order 15 per cent towards higher halo abundances, i.e. towards the PS form. The median mass increase for individual dark matter haloes is of order 25 per cent for galaxy-sized objects and of order 20 per cent for clusters. However there is still a considerable gap between our corrected mass function (Eq. \ref{fittingfunction}) and the PS mass function (Fig. \ref{massfunction}). This can be partly understood by again tracing back all particles, that form the total mass, to their initial positions. Many of the disconnected FoF regions (see Section \ref{code}) are now connected, but not all of them. The typical initial region which will form part of the total mass of a $z=0$ halo still differs  form a simple spherical or ellipsoidal overdensity, even though the difference does decrease when the total mass is used instead of the FoF or SO mass. Furthermore since the total mass correction is more or less uniform over the entire mass range, we find a  higher abundance of clusters and superclusters than predicted by PS. This suggests that the PS ansatz might be too simple to allow precise predictions about abundance and properties of dark matter haloes.
\newline
The classical secondary infall model \citep[][]{Fillmore,Bertschinger} describes the median orbits of particles and subhalos in the outer regions of halos very accurately \citep[][]{Diemand2}. Extensions of this model match the mass distributions in and around simulated halos \citep[][]{Ascasibar}. It could therefore be feasible to calculate the expected halo mass fraction beyond the virial radius within this framework and it would be useful to compare such results with our measured median corrections (Fig. 3).
\newline
\indent \citet[][]{Robertson} used cosmological simulations to test the excursion set ansatz by identifying the locations in the linear overdensity field that later collapse to form dark matter haloes. They found an inconsistency between the effective collapse barrier of simulated haloes and excursion set formalism predictions for their abundance, and conclude that the excursion set ansatz fails, i.e. that the extended PS formalism cannot predict halo abundances exactly. However only the common mass definitions (FoF and SO) were considered and it would be worthwhile to reinvestigate these issues using the total collapsed mass instead. Using the total mass will also allow to derive physically meaningful mass accretion and merger rates from cosmological simulations, and it will change halo formation times and their spatial clustering.
\newline
\indent Taking the total mass of a halo into account, instead of adopting an artificial truncation at the virial radius, does make a difference in the interpretation of the kinematics of groups and clusters. For example, it increases escape velocities significantly. The fast neighbours AndXIV \citep[][]{Majewski} and AndXII \citep[][]{Chapman} might well be bound to Andromeda without requiring a much larger {\it virial} mass. And even objects which are quite isolated today, e.g. Tucana \citep[][]{Mateo}, might well have had a close interaction with their primary (in this case the Milky Way) at some earlier time and might have another one in the future.

\section*{Acknowledgments}
We thank Darren Reed, Antonio J. Cuesta and Marcel Zemp for helpful discussions as well as Doug Potter and Alejandro Daleo for support in the development of the code. Further we would like to thank the anonymous referee for useful comments. This work was supported by the SNF.

\label{lastpage}
 
\end{document}